# The Effects of Technology and Innovation on Society

Peter Sasvari

**Abstract- Various models of the information society have been developed so far and they are so different from country to country that it would be rather unwise to look for a single, all-encompassing definition. In our time a number of profound socio-economic changes are underway. The application of these theories and schools on ICT is problematic in many respects. First, as we stated above, there is so not a single, widely used paradigm which has synthesised the various schools and theories dealing with technology and society. Second, these fragmented approaches do not have a fully-fledged mode of application to the relationship of ICT and (information) society. Third, SCOT, ANT, the evolutionary- or the systems approach to the history of technology when dealing with information society – does not take into account the results of approaches studying the very essence of the information age: information, communication and knowledge. The list of unnoticed or partially incorporated sciences, which focuses on the role of ICT in human information processing and other cognitive activities, is much longer.**

## I. INTRODUCTION

Many theories can be found in the literature on the information society. The theories of the knowledge or information economy, postindustrial society, postmodern society, information society, network society, information capitalism, network capitalism etc. show that it is an important sociological issue to understand what role is played by technology and information in the society we live in. Both aspects - the form of society and the role of technology and information - belong to the central question of the theory of the information society [26] [27].

## II. THE DEFINITION OF INFORMATION

In everyday use, the term "information" has meant a kind of guidance for a long time: when someone goes to the railway station to be informed on the content of the timetable or to the information desk to find out where a product or a counter can be found in the department store. Such information exchange works only if the right piece of information, the one that fits and makes sense for both parties of the communication is available.

Peter Sasvari, Institute of Business Sciences, Faculty of Economics, University of Miskolc, Hungary, 3515, Miskolc-Egyetemvaros, E-mail: iitsasi@uni-miskolc.hu. The described work was carried out as part of the TÁMOP-4.2.1.B-10/2/KONV-2010-0001 project in the framework of the New Hungarian Development Plan. The realization of this project is supported by the European Union, co-financed by the European Social Fund. Manuscript received February 26, 2012; revised March 30, 2012 and November 15, 2012

Information as a term became more and more popular in the last 30-40 years; it has started to have an increasingly important role in everyday language while its strict meaning mentioned above has gradually faded away. At the same time, there has been a growing uncertainty about the true meaning of the term 'information'. All this doubtfulness is mainly caused by the so-called 'information-centered' world we are living in and by the widespread expansion of information and communication technology as almost everyone living in developed western societies can experience the phenomenon called the Information Revolution. All this suggests that information has become an essential part of our society and plays a centre role in our lives. In information studies, rather complex definitions can be found on the nature of information. Informatics or computer science is a discipline that deals with the storage, processing and distribution of information as well as planning computer networks and determining their operation principles. Determining the exact subject of computer science is rather difficult because it is extremely hard to define what information is. According to the German physicist and philosopher, Weizsacker information should be regarded as the third universal elementary quantity beside matter and energy in science and technology [32].

According to the ninth volume of the Pallas Nagy Lexikon (Great Pallas Encyclopedia) information is a term with Latin origins meaning report, enlightment, inform, let somebody know; informant, instructor, messenger [15].

In the Dictionary of Foreign Words and Expressions the following meanings can be read:

1. Enlightment, announcement, communication;

2. Message, data, news, bulletin.

The fourth edition of American Heritage Dictionary of the English Language distinguishes seven meanings of the term 'information':

1. Knowledge derived from study, experience, or instruction.

2. Knowledge of specific events or situations that has been gathered or received by communication; intelligence or news.

3. A collection of facts or data: statistical information

4. The act of informing or the condition of being informed; communication of knowledge.

5. Computer Science Processed, stored, or transmitted data.

 

6. A numerical measure of the uncertainty of an experimental outcome.

7. Law a formal accusation of a crime made by a public officer rather than by grand jury indictment.

The theory of communication states that information is the objective content of the communication between objects conversely affecting each other which is manifested in the change of the condition between these objects.

According to the science of telecommunication information is a series of signals structured in time and space, which is made up of a signal set having a specific statistical structure. The sender transmits data on the condition of an object or on the course of an event and the receiver perceives and interprets these signals. Everything can be regarded as information that is encoded and transmitted through a definite channel.

From the perspective of social science, information is the communication of useful knowledge that is created and transmitted in the intellectual communication system of the society. It is characteristic to the society as a whole, belonging to one of the global issues of world together with energy and environment protection.

According to the economic approach, information is partly a form of service, partly a product but, not as in the case of exchange of goods, both parties can keep their information. The content of material, energy and living labour is gradually decreasing in manufactured goods, while the amount of product information input is increasing at the same rate.

In summary, information is an expression related to enlightment, data, report, learning, communication and news. In certain cases, it can be identified with these items (knowledge, data, enlightenment, news); in other cases it is the object of these listed items (conveyance of knowledge, learning, communication).

Despite the fact that it may still sound uncertain, the group of the terms 'data', 'knowledge' and 'communication' can be highlighted for giving an interpretation of information. According to the literature, the transformation of data into information needs knowledge. There are many definitions trying to find a link between information and communication, which also can have an importance when looking for a definition of the information society. Communication is a process of transferring information from one entity to another through a specific medium. If we link these two different approaches together, the picture we are given is a very complex one, where the four terms 'data', 'information', 'knowledge' and 'communication' must be interpreted in one compound definition. The same connection was made by Michael Buckland in his book on information systems.

TABLE I
FOUR ASPECTS OF INFORMATION [3]

|  | Intangible | Tangible |
|---|---|---|
| Entity | Information as knowledge | Information as thing |
| Knowledge | Information as process | Information process |

Information as knowledge is subjective in every case, it is linked to a given individual and it gains its exact meaning in a specific environment. It is intangible as an entity but it can be communicated, made to be known to others. Information as a thing exists similarly to knowledge, however, it is tangible. In this regard, data can be regarded as a kind of recorded knowledge because it is necessary to know the context of its creation (or the record structure), without having this context, the data cannot be interpreted.

## III. INFORMATION AND SOCIETY

By definition, society refers to:

1. Human relations and relationships taken as a whole,

2. Any community of human beings is able to perpetuate itself, more or less linked to a specific region or country sharing a distinctive culture and institutions.

Whether a human community is regarded as a society depends on the extent to which its members are able to interact with each other, thus the capacity and extension of interaction is essential.

The most recent trends show that the definition of society has become less important in trying to understand the world surrounding us because if we examine only individual societies, we may not notice social (multilateral and global) phenomena between and over societies.

If we accept that the key feature of social existence is the development of relations, then the information society may bring a significant change in this very context: a lot more individuals have the opportunity to get in contact with other people in a simpler way and at a lower cost.

A question comes up here immediately: is it possible to call every human society an information society? Information is the essential condition of the functioning of every society, including their subsystems as well. It played an important role in every social formation in the agricultural and industrial societies of previous ages. Information flow is needed in every society but none of the previous societies were labelled "information society" by contemporary analysts and historians. The reason for this is that the communication, reception, processing, storage, interpretation and flow of information never determined earlier societies to such a high extent as today's. The activities relating to information have become more valuable in present day societies and that is what distinguishes them sharply from the societies of the past. This fundamental difference is convincingly described by theoretician of various interests, views and attitudes and orientation in the following five fields:

1. Technology.
2. Occupation structure.
3. The operation of economy.
4. Spatial structure.
5. Culture.

Frank Webster's book published in 1995 synthesizes the 1960s and 1970s information society theories in order to analyze the concept and its characteristics within the context of social science [30]. These theories designate the potential directions of what might be a comprehensive research project, which can clarify the concept and exploit these theories as starting points for further exploration. Webster's typology is the following [31]:

### A. Technology

From the technological perspective we live in an information society since information and telecommunication technologies play a constantly expanding role in all fields of social existence, which has shaken the foundations of social structures and processes and resulted in profound changes in politics, economy, culture, and everyday life.

Most of the attempts made to define information society approach the idea from a technological point of view hence the central question of such explorations sounds like: what kind of new information and communication technology was constructed in recent decades that determined the infrastructure of information society?

### B. Occupation structure and economy

Studies of occupational structure and economy show that we live in an information society because, when we have passed through the agricultural and industrial stages, the information sector and information oriented jobs dominate the economy. The main questions raised by this approach are: How have the proportions of employed workers changed in the industrial and service sectors in recent decades? How have their performance and the knowledge they use changed qualitatively? Have the so-called informational occupations begun to dominate production?

The question is similar to that which we posed by the technological approach: What is the point at which we can claim that the logic of capitalism, that is, its structure of production has qualitatively changed? Is the often cited "new economy" indeed so different from the old one? Where is the turning point? Is it possible to identify the point at which the former was replaced by the latter?

### C. Spatial structure

As the spatial theorists see it we live in an information society because through the use of information technologies and globalization physical space tends to lose its determining function. People are participating in networks that determine such social processes as production, division of labour, discussing politics for example.

The main theoretical questions are the following: Does the world follow the logic of networks? Does global society exist? Can it come to life? What is the inherent logic of global networks? Who belongs to them, and why do they wish to do so? What kind of social and economic capital is needed to gain access to a network and how can membership then be maintained? What are the innate social relations of the network, and what part do the new information and communication technologies play in those relations?

### D. Culture

The cultural perspective also states that we live in an information society because our life is infiltrated by the globalised, extensively digitalized media culture that has become the primary means of providing sense and meaning for us and predominantly determines our lifestyle.

Theories attempting to explain the cultural aspects of information society describe such a global cultural context that may be adopted universally as a referential framework for the media. This approach also suggests that the media enjoy a unique status in the age of information and that they are the most prominent determining factors of social relations.

However, the question remains: whether life exists beyond media culture or not? Does the illusory game of signs have any connection to reality? The catchphrase of the information age is "virtual reality" which reality very often turns out to be more fundamental than the world that created it.

## IV. THE INFORMATION SOCIETY

Many theories can be found in the literature on the information society. The theories of the knowledge or information economy, postindustrial society, postmodern society, information society, network society, information capitalism, network capitalism etc. show that it is an important sociological issue to understand what role is played by technology and information in the society we live in. Both aspects - the form of society and the role of technology and information - belong to the central question of the theory of the information society.

One of the first social scientist to develop the concept of the information society was the economist Fritz Machlup [13]. In his breakthrough study, "The production and distribution of knowledge in the United States" (1962), he introduced the concept of the knowledge industry by distinguishing five sectors of the knowledge sector:

- Education.
- Research and development.
- Mass media.
- Information technologies.
- Information services.

Peter Drucker has argued that there is a transition from an economy based on material goods to one based on knowledge [7] [8] [9].

Marc Porat distinguishes [17] [18]

- A primary sector (information goods and services that are directly used in the production, distribution or processing of information) and

- A secondary sector (information services produced for internal consumption by government and non-information firms) of the information economy.

Porat uses the total value added by the primary and secondary information sector to the GNP as an indicator for the information economy. The OECD has employed Porat's definition for calculating the share of the information economy in the total economy. Based on such indicators the information society has been defined as a society where more than half of the GNP is produced and more than half of the employees are active in the information economy.

For Daniel Bell the number of employees producing services and information is an indicator for the informational character of a society [1] [2]. A post-industrial society is based on services. What counts is not raw muscle power, or energy, but information. A post industrial society is one in which the majority of those employed are not involved in the production of tangible goods.

1. Economic sector.
2. Resource.
3. Strategic resource.
4. Technology.
5. Knowledge base.
6. Methodology.
7. Time perspective.
8. Planning.
9. Guiding principle.

TABLE II
DIMENSIONS OF THE INFORMATION SOCIETY [2]

| | PREINDUSTRIAL SOCIETY | INDUSTRIAL SOCIETY | POST INDUSTRIAL SOCIETY |
|---|---|---|---|
| MODE OF PRODUCTION | Extractive | Fabricating | Processing; Recycling |
| ECONOMIC SECTORS | Primary: agriculture, mining, fishing, timber | Secondary: manufacturing, processing | Tertiary: transportation, utilities, Quaternary: trade, finance, insurance, real estate, Quandary: health, education, research, government, recreation |
| TRANSFORMING RESOURCE | Natural power: wind, water, draft animals, human muscle power | Created energy: Electricity, oil, gas, coal, nuclear power | Information Computer and data transmission systems |
| STRATEGIC RESOURCES | Raw materials | Financial capital | Knowledge |
| TECHNOLOGY | Craft | Machine technology | Intellectual technology |
| KEY OCCUPATIONS | Farmer, miner, fisherman, unskilled worker | Semi-skilled worker, engineer | Professional and technical occupations, scientists |
| KEY METHODS | Common sense, trial and error, practice | Empiricism, experimentation | Abstract theories, models, simulations, decision theory, system analysis |
| TIME PERSPECTIVE | Orientation to the past | Ad hoc adaptiveness, experimentation | Future-oriented prediction and planning |
| DESIGN | Game against nature | Game against fabricated future | Game between individuals |
| GUIDING PRINCIPLE | Traditionalism | Economic growth | Codification of theoretical knowledge |

Alain Touraine already spoke in 1971 of the post-industrial society [28]. "The passage to postindustrial society takes place when investment results in the production of symbolic goods that modify values, needs, representations, far more than in the production of material goods or even of 'services'. Industrial society had transformed the means of production: post-industrial society changes the ends of production, that is, culture. The decisive point here is that in postindustrial society all of the economic system is the object of intervention of society upon itself. That is why we can call it the programmed society, because this phrase captures its capacity to create models of management, production, organization, distribution, and consumption, so that such a society appears, at all its functional levels, as the product of an action exercised by the society itself, and not as the outcome of natural laws or cultural specificities".

In the programmed society also the area of cultural reproduction including aspects such as information, consumption, health, research, education would be industrialized. That modern society is increasing its capacity to act upon itself means for Touraine that society is reinvesting ever larger parts of production and so produces and transforms itself. This idea is an early formulation of the notion of capitalism as self-referential economy.

In Yoneji Masuda's framework, industrial and information societies are compared to one another by 20 different indicators [14].

Similarly to Bell Peter Otto and Philipp Sonntag assert that an information society is a society where the majority of employees work in information jobs, i.e. they have to deal more with information, signals, symbols, and images than with energy and matter.

Radovan Richta argues that society has been transformed into a scientific civilization based on services, education, and creative activities. This transformation would be the result of a scientific-technological transformation based on technological progress and the increasing importance of computer technology. Science and technology would become immediate forces of production [19].

TABLE III
COMPARISON OF THE CHARACTERISTICS OF THE INDUSTRIAL AND
INFORMATION SOCIETY

| | | INDUSTRIAL SOCIETY | INFORMATION SOCIETY |
|---|---|---|---|
| INNOVATIONAL TECHNOLOGY | Core | Steam engine (power) | Computer (memory, computation, control) |
| | Basic function | Replacement, amplification of physical labour | Replacement, amplification of mental labour |
| | Productive power | Material productive power (increase in per capita production) | Information productive power (increase in optimal action-selection capabilities) |
| SOCIOECONOMIC STRUCTURE | Products | Useful goods and services | Information, technology, knowledge |
| | Production centre | Modern factory (machinery, equipment) | Information utility (information networks, data banks) |
| | Market | New world, colonies, consumer purchasing power | Increase in knowledge frontiers, information space |
| | Leading industries | Manufacturing industries (machinery industry, chemical industry) | Intellectual industries (information industry, knowledge industry) |
| | Industrial structure | Primary, secondary, tertiary industries | Matrix industrial structure (primary, secondary, tertiary, quaternary systems/industries) |
| | Economic structure | Commodity economy (division of labour, separation of production and consumption) | Synergetic economy (joint production and shared utilization) |
| | Socioeconomic principle | Law of price (equilibrium of supply and demand) | Law of goals (principle of synergetic feed forward) |
| | Socioeconomic subject | Enterprise (private enterprise, public enterprise, third sector) | Voluntary communities (local and informational communities) |
| | Socioeconomic system | Private ownership of capital, free competition, profit maximization | Infrastructure, principles of synergy, precedence of social benefit |
| | Form of society | Class society (centralized power, classes, control) | Functional society (multicentre, function, autonomy) |
| | National goal | GNW (gross national welfare) | GNS (gross national satisfaction) |
| VALUES | Form of government | Parliamentary democracy | Participatory democracy |
| | Force of social change | Labour movements, strikes | Citizens' movements, litigation |
| | Social problems | Unemployment, war, fascism | Future shock, terror, invasion of privacy |
| | Most advanced stage | High mass consumption | High mass knowledge creation |
| | Value standards | Material values (satisfaction of physiological needs) | Time-value (satisfaction of goal achievement needs) |
| | Ethical standards | Fundamental human rights, humanity | Self-discipline, social contribution |
| | Spirit of the times | Renaissance (human liberation) | Globalism (symbiosis of man and nature) |

Nico Stehr says that in the knowledge society a majority of jobs involves working with knowledge. "Contemporary society may be described as a knowledge society based on the extensive penetration of all its spheres of life and institutions by scientific and technological knowledge" [21] [22] [23] [24].

For Stehr knowledge is a capacity for social action. Science would become an immediate productive force, knowledge would no longer be primarily embodied in machines, but already appropriated nature that represents knowledge would be rearranged according to certain designs and programs. The economy of a knowledge society is largely driven not by material inputs, but by symbolic or knowledge-based inputs, there would be a large number of professions that involve working with knowledge, and a declining number of jobs that demand low cognitive skills as well as in manufacturing.

Also Alvin Toffler argues that knowledge is the central resource in the economy of the information society: "In a Third Wave economy, the central resource – a single word broadly encompassing data, information, images, symbols, culture, ideology, and values – is actionable knowledge".

In recent years the concept of the network society has gained importance in information society theory. For Manuel Castells network logic is besides information, pervasiveness, flexibility, and convergence a central feature of the information technology paradigm. "One of the key features of informational society is the networking logic of its basic structure, which explains the use of the concept of 'network society'" As a historical trend, dominant functions and processes in the Information Age are increasingly organized around networks. Networks constitute the new social morphology of our societies, and the diffusion of networking logic substantially modifies the operation and outcomes in processes of production, experience, power, and culture. For Castells the network society is the result of informationalism, a new technological paradigm [4] [5].

Jan Van Dijk defines the network society as a "social formation with an infrastructure of social and media networks enabling its prime mode of organization at all levels (individual, group/organizational and societal) [6]. Increasingly, these networks link all units or parts of this formation (individuals, groups and organizations)". According to Van Dijk networks have become the nervous system of society, whereas Castells links the concept of the network society to capitalist transformation, Van Dijk sees it as the logical result of the increasing widening and thickening of networks in nature and society.

The major critique of concepts such as information society, knowledge society, network society, postmodern society, postindustrial society, etc. that has mainly been voiced by critical scholars is that they create the impression that we have entered a completely new type of society. If there is just more information then it is hard to understand why anyone should suggest that we have before us something radically new.

Such neomarxist critics as Frank Webster argue that these approaches stress discontinuity, as if contemporary society had nothing in common with society as it was 100 or 150 years ago [30]. Such assumptions would have ideological character because they would fit with the view that we can do nothing about change and have to adopt to existing political realities. These critics argue that contemporary society first of all is still a capitalist society oriented towards accumulating economic, political, and cultural capital. They acknowledge that information society theories stress some important new qualities of society (notably globalization and informatization), but charge that they fail to show that these are attributes of overall capitalist structures. If there were a discourse on continuity and discontinuity, capitalism would enter into a new development stage.

Concepts such as knowledge society, information society, network society, informational capitalism, postindustrial society, transnational network capitalism, postmodern society, etc. show that there is a vivid discussion in contemporary sociology on the character of contemporary society and the role that technologies, information, communication, and co-operation play in it. Information society theory discusses the role of information and information technology in society, the question which key concepts shall be used for characterizing contemporary society, and how to define such concepts. It has become a specific branch of contemporary sociology.

## V. MAKING THE INFORMATION SOCIETY QUANTIFIABLE

At the end of the overview of examination criteria comes a synthetic table which partly improves the previous models and partly specifies them. This table includes formulations to make individual elements measurable and thus answers the question of from which point and to what extent of deviation from absolute or relative indicators can a society be regarded as an information society. That is, where is the tipping point from one state to another in a sub-system or in regard to a characteristic, and through this, of all society? The same table will demonstrate that in many cases it is typical of metaphors found in book titles to focus only on particular limited areas. Returning to the idea proposed in the introduction, we should restate that the term information society is not a "rival" of these terms but an umbrella term incorporating them all.

TABLE IV
SYNTHETIC BASIC CATEGORIES OF INFORMATION SOCIETY, THEIR MEASURABILITY AND METAPHORS [11] [12]

| BASIC CATEGORY | MEASURE AND "TIPPING POINT" | METAPHOR |
|---|---|---|
| PRODUCTION (MANUFACTURING) | The proportion of businesses forming part of the information sector and producing information and knowledge products in relation to other sectors (relative dominance: when it is the largest sector; absolute dominance: when the sector alone produces over 50%, i.e. it is larger than all the others put together). | Information industry, knowledge industry, information and knowledge industry, information economy, knowledge economy, knowledge-based economy |
| EMPLOYMENT | The number and proportion of those employed in the information and knowledge sectors in relation to other sectors (relative dominance: when it is the largest sector; absolute dominance: when the sector alone produces over 50%, i.e. it is larger than all the others put together). | White-collar workers, information and knowledge workers, immaterial workers, knowledge class intelligentsia |
| WORK | How many people and to what degree are engaged in information activity "as a profession" according to the type of work done (threshold level: 50%). | Symbol manipulators, intelligence, brainworker/mind worker |
| RESOURCE AND TECHNOLOGY | Information and knowledge appear as resources and forms of capital in addition to traditional forms – the theory of growth and accounting strive to mathematise this but so far there are no accepted algorithms. (However, the contribution of information and knowledge technology to growth is already measured.) | Intellectual capital, human capital, information capital, corporate information and knowledge assets |
| INCOME AND WEALTH | GNP at a national level, monthly income on an individual level. There are no accepted measures in regard to the amounts; what is more, these amounts vary depending on the time of joining the information society. $5,000/person/month was the threshold level at the turn of the | Affluence, welfare state |
| | 1960s in the USA. | |
| CONSUMPTION | The proportion of purchased information and cultural goods, means and services in the consumer basket, especially in regard to media contents (threshold level: 33%). | Consumer society, prosumers, mediatised society |
| EDUCATION (LEVEL OF EDUCATION) | Proportion of those with a qualification earned in higher education (degree holders) in society (threshold level: 50%). | Learning society, meritocracy |
| COGNITION | Results and scales in the measurable dimensions of cognition; microscopic dimensions, astronomical distances and scales, discovered genocombinations, sign processing, etc. The scale to measure this is still to be worked out. | Life-long learning, scientific revolution, nano-scale, peta-scale |
| CONFLICT MANAGEMENT METHOD AND POWER TECHNIQUE | Replacement of traditional forms of warfare, placing economic conflicts into an information context (business intelligence, innovation competition). The "state of democracy" of society, types and mediators of control. There are some methods used to measure the "degree" of democracy. | Information warfare, cyber wars, business intelligence, bureaucracy, control crisis- and revolution, risk society |
| INTER-CONNECTEDNESS | The degree of mutual connectedness (objective in the case of telephone networks: provision over 50%). | Telematic society, "wired society" |
| WORLDVIEW AND LOGICAL FRAMEWORK | Has the static and energy-centred worldview been replaced by an information-centred one? Have the global system level and the "space age" become a framework for analysis and interpretation? Is orientation to the future a characteristic feature? | Global village, techno culture, information civilisation |

## VI.  LEGAL REGULATIONS OF INFORMATION SOCIETY

The legal material concerning information society is interwoven into our legal system horizontally. The rules related to information society are enshrined to a greater or lesser extent in the several areas of law. As in any regulatory domain, the legal content concerning information society can be grouped according to the system of law. There are two distinct groups: the laws organizing legal relations between the state and its citizens, and between the various state or public organizations (called public law), and the laws organizing legal relations between citizens and partnerships, and between members of civil society (civil law) [10]. Differentiation is based on the relationship between those involved. While in the first case we can speak of an unequal legal relation based on subordination and superiority, in civil law the typical legal relation is one of equality and coordination.

In the continental legal system, we can distinguish between four main categories:

1. Civil law.
2. Criminal law.
3. Administrative law.
4. Constitutional law.

Civil law regulates the property personal and family relations of natural and legal persons in cases where the partners are equal and state intervention, except for legislation, occurs only in the event of a legal dispute. The most important areas affecting information society are as follows:

- E-commerce.
- Digital signature.
- Content regulation.
- Protection of copyright and industrial property rights,
- Media law.
- Competition law.

Criminal law regulates acts that are a danger to society. We can group all those acts committed with or against IT technology which are dangerous for society and for which the law orders the sanction of punishment. Legal regulation of information society is primarily concerned with the following categories of crime:

- Misuse of personal data.
- Content-related crimes (e.g. distribution of child pornography hate speech, etc.).
- Crimes against computer systems and data.
- Infringement of copyright.

Administrative law is the regulatory system of state functions. State administration extends beyond central government and local government to larger systems; for example the operation of transport, security, military and information systems. The following functions essential to information society belong to this group:

- Electronic administration.
- Electronic register of companies.
- Administrative procedure.
- Electronic public procurement.

The fourth field is constitutional law, which arose out of continental legal development. The object of regulation is to structure relations between the citizens and the state and the organizational structure of the state. The constitution is the document describing basic rights, responsibilities and procedures thus creating the basis for the process governing political, economic and social life. Areas of constitutional law related to the information society are as follows:

- Electronic freedom of information.
- Personal data protection.
- Freedom of the press and freedom of expression.

## VII. THE EFFECTS OF TECHNOLOGY AND INNOVATION ON SOCIETY

Technique can be defined as the application of some devices or knowledge in order to accomplish a specific task or fulfill a purpose. These purposes may range from industrial use to social needs, improving working conditions or raising the standard of living. For humans, technique is an acquired way of using the surrounding environment for satisfying their own instinctive goals and cultural desires. It is the knowledge to create something new.

Under the term 'technology' I mean all the procedures and knowledge of procedures that are needed to perform a specific task.

Studies considering science and technology as an inseparable and organic part of society, like information society studies, do not have a unified conception and methodological apparatus, nor a comprehensive and prevailing scientific paradigm. We can talk about a variety of multidisciplinary and interdisciplinary studies, schools, theories and approaches interacting with each other and comprising works of scholars from various traditional sciences like history, economics, sociology or anthropology. The great number of diverse approaches makes it impossible to review them completely, so we have to forget about introducing schools like the technology theories of evolutionary economics in detail. On the whole, the goal of this chapter can be nothing more than to provide an "intellectual crutch" for discussing and interpreting information communication technologies by reviewing the most relevant and important theories, concepts, models and notions of the topic.

Technological determinism argues that technology is the principal driving force of society determining its mode of operation, development, course of history, structure and values in a decisive manner. Converse effects are taken into account to a limited extent, fully disregarded or disclaimed. Technological development is thought to be propelled by the logic of science alone.

Most scientific concepts explicitly reject technological determinism; yet they assist its survival by studying only technology's influence on society. This is more symptomatic of ICT related researches.

The beginning of Science, Technology and Society studies dates back to the early 1970s, when the first studies were published. The novelty in the pioneering works, which lends them their special character even today, was that they stressed –contrary to technological determinism – society's crucial role in the development of science and technology, framing the three intermingling domains in complex theoretical systems. The works of philosophers, historians and sociologists were collected in two books in the mid-eighties, which have become the most cited publications of this school. Some of these approaches have developed into theories, generating further discourses and STS has been crystallized into an interdisciplinary field of research with both common research areas and methodology.

The STS school is far from being the dominant scientific paradigm of this area of knowledge, but has several advantages that make it indispensable when examining information society and ICT. These are its strong empirical basis and complex approach to analyzing interaction between technology and society, their manifold co-dependence, and complex co-development. Within the several concepts of STS, many schools exist criticizing and complementing each other.

### A. Studies of the interactions between science, technology and society

The foundations of STS were laid down in the 1980s by the "Social Construction of Technology" school, which focuses on the development phase of technologies at the micro level, and pinpoints that technology (and natural scientific developments) are basically shaped by social processes.

Any given technology stabilizes when debates are settled. This is the phase of 'closure and stabilization'. Closure, however, does not mean finalizing: newly joined user groups can reopen the debates which can lead to new modifications to or variations of the existing technology.

Using the terminology of evolutionary approaches, we can say variations, mutations and hybrids are brought to life during the diffusion of a certain technology, which is chiefly true for ICT. Take the different variations of computers (desktop PC, portable notebook, PDA, etc.) or the convergence of mobile phones with other electronic devices (such as PDAs, digital cameras, mp3-players, game consoles, or GPS devices) which are typical hybrids.

Bijker and Pinch emphasize that the meanings assigned to technologies are determined by the norms and values of social groups which draw the "wider context" of socio-cultural and political environment into the set of determining factors. Drawing on the wider context concept, Laudan R. argues that changing social values can bring new technological constructs or their complete generation to life. The heterogeneous and hierarchical community of technological development functions as a mediator of social values and forces value orientation in society to change.

Capital mobility has increased incredibly, the economy has shifted to the service sector, innovation has become the primary source of productivity growth in relation to engineering, organizations, institutions as well as individual workers. The "technical construction of society" has become a major issue, that is social processes are mainly mediated by technological development. The society's level of being informed, with exploiting the opportunities provided by information and communication technology, has been increased dramatically. This new technology, together with biotechnology opens new perspectives in the fields of industry, trade and education.

The nature of economic competition has been undergoing huge changes, as more and more people think that there have been profound changes in the relation between economy and society and innovation requirements. The continuous and self-accelerating innovation processes characterized by the intense competition has brought about some changes in time relations. People start moving on a different time scale, time has been speeded up.

Space has become globalized, by turning into more unified and more complex at the same time. Socio-economic processes create new virtual spaces or even real spaces are modified: the processes are arranged in new ways in the interacting local, regional, national and supranational spaces. While integration processes are considered to be a general tendency, clear attempts for isolation also appear repeatedly. Knowledge has become the main economic source, and learning abilities and skills have become a criterion of adaptation at the levels of individuals, companies, local communities, nations, supranational organizations and the world taken as a global system.

Actor-Network-Theory is another school of STS studies, which is more and more widely used. It is a new branch of the sociology of science and technology, the basis of which was elaborated by Michel Callon, Bruno Latour and John Law in the 1980s. They – along with other scholars – developed their concepts into a theory.

A basic statement of ANT is that technological objects along with their socio-political context co-develop and shape each other mutually into socio-technical entities through constant interactions. The objects and their context form heterogeneous networks made up of human and non-human components which are connected to each other dynamically. These heterogeneous components can be objects, techniques, institutions, organizational solutions, human abilities or cognitive structures.

Human components as network builders are constantly formed and constituted by the networks they are part of. Actors in this network are connected by intermediaries, which in many cases, have social meanings. Texts, technical artefacts, currencies or human skills can function as intermediaries.

One of ANT's – much debated – theorems is that the natural state of society is disorder. Order is achieved through the constant and endless efforts made by the actors to build networks.

Callon argues that an actor-network cannot be derived either from the actor or the network. The actions and the will of actors are inseparable from the network, and their effect runs through the whole network.

This leads us to one of ANT's radical novelties: the boundaries between the actors disappear and even actions cannot be interpreted in the traditional way.

In the literature, the constant shifting of power between technology and society is called translation: as a result of this process, networks are formed progressively, in which certain entities gain control over other entities.

### B. Diffusion of innovations

Innovation has become a key activity of information societies. It is the cornerstone of economic competitiveness. National and regional (such as European) administrations develop high level strategies to promote innovative activities in the economy.

Innovation can be defined as basically novel inventions or concepts – arising from either professional research or ideas by amateurs – translated into practice. An innovation can be a technological object, a new organizational solution or an idea. Innovations become market goods through product development and/or technology transfer. The product cycle consists of the following stages: introduction (to the market), growth, maturity and stabilization, and decline. The life cycle of common goods (e.g. road infrastructure) and public goods (e.g. public safety) go through the same stages. Rogers' theory applies to the life cycle of innovations as far as the maturity phase and at the level of communities and societies.

Rogers explains the diffusion of innovations as basically communicative: diffusion is the process by which an innovation is communicated through certain channels over time among the members of a social system. Diffusion is determined by the above mentioned four factors (innovation, communication channels, time and social systems). It is a process of decision making, in the stages of which different types of information and knowledge transferring mechanisms play crucial roles.

The diffusion of innovations – thus, of technologies too – takes place within social networks, so called diffusion networks. The ability of individuals to adapt depends on the cohesion of these networks, in other words, to the extent of its homophily (similar socio-economic status, qualifications, attitudes); on structural equivalence (on the individual's position in the network); and on the threshold of other users which makes it worthwhile for a group member to adopt the given technology.

Innovators play a crucial role in diffusing an innovation between homophile diffusion networks. They tend to use the technology first, and usually possess heterophile social relations (they maintain regular relationships with several social groups and through them, several networks of diffusion). Chronologically, the second group to adopt an innovation are called the early adopters; these are followed by the early majority, then the late majority, and lastly, the laggards. Each of these ideal-typical groups is characterized by specific socioeconomic factors, personality values and communication behavior. For example laggards are the most disadvantaged group along the socio-economic scale.

When studying the diffusion of ICT, at least one more category must be added: the refusers, who consciously resist usage throughout their lives (also known as diehards). The existence of this group indicates that no technology ever penetrates a society fully. To reach 100% diffusion both society and technology need to change as compared to their initial status when the innovation was introduced.

The process of diffusion is broken down into different stages from the individual user's point of view. First, one typically acquires information regarding innovation through mass media channels (or cosmopolitan communication channels). The following three phases are dominated by interpersonal channels (or local channels). In the second phase, persuasion and opinion forming take place, followed by deciding on the adaptation, finally evaluation and confirmation of the usage. Of course, refusing the implementation (even several times) is an option too, but it can be followed by acceptance, and vice versa, the evaluation of implementation can lead to discontinuing usage. Rogers analyses the characteristics of an innovation affecting its own diffusion (such as relative advantage, compatibility, complexity, trial ability and observe ability), but gives little attention to their socially constructed nature [20].

The main advantage of Rogers' theory is that a key role is ascribed to communicative processes. This momentum makes the theory a close relative to other approaches such as SCOT and ANT. Rogers' theory can be drawn upon in the analyses of such information society related issues as the digital divide or e-inclusion.

## VIII. CONCLUSION

The application of these theories and schools on ICT is problematic in many respects. First, as we stated above, there is not a single, widely used paradigm, which has synthesized the various schools and theories dealing with technology and society. Second, these fragmented approaches do not have a fully-fledged mode of application to the relationship of Information Control Technology (ICT) and (information) society [25].

Third, SCOT, ANT, the evolutionary- or the systems approach to the history of technology – when dealing with information society – does not take into account the results of approaches (such as information science or information systems literature or social informatics, information management and knowledge management, communication and media studies) studying the very essence of the information age: information, communication and knowledge. The list of unnoticed or partially incorporated sciences, which focuses on the role of ICT in human information processing and other cognitive activities, is much longer [29].

These, though, miss the approach of STS and evolutionary schools, particularly the concept of technology and society as a seamless web. Merging the two modes of understanding information society is in its infancy, though studying ICT systems cannot be completes without them both [27].